\documentclass[a4paper,12pt]{article}
\usepackage[latin1]{inputenc}
\usepackage[T1]{fontenc}
\usepackage{sectsty}
\usepackage{setspace}
\usepackage{graphicx}
\usepackage{amsmath, amsthm, amssymb}
\usepackage{enumerate}
\usepackage{enumitem}
\usepackage[margin=20mm]{geometry}
\usepackage{natbib}
\usepackage[english]{babel}
\usepackage{hyperref}
\usepackage{authblk}
\usepackage{lmodern}
\usepackage{csquotes}
\usepackage{setspace}
\usepackage{latexsym}
\usepackage{mathptmx}
\usepackage{braket}
\usepackage{txfonts}
\usepackage{cleveref}
\usepackage[table,usenames,dvipsnames]{xcolor}
\usepackage{comment}
\definecolor{darkblue}{rgb}{0.0,0.0,0.3}
\hypersetup{colorlinks,breaklinks,linkcolor=darkblue,urlcolor=darkblue,anchorcolor=darkblue,citecolor=darkblue}

\def\citepos#1{\citeauthor{#1}'s (\citeyear{#1})}

\usepackage{ifthen}
\def\eprinttmp@#1arXiv:#2 [#3]#4@{\ifthenelse{\equal{#3}{}}{\href{http://arxiv.org/abs/#1}{arXiv:#1}}{\href{http://arxiv.org/abs/#2}{arXiv:#2 [#3]}}}
\newcommand{\eprint}[1]{\eprinttmp@#1arXiv: []@}
\newcommand{\doi}[1]{\href{http://dx.doi.org/#1}{doi:#1}}

\sectionfont{\normalfont\large\bfseries}
\subsectionfont{\normalfont\normalsize\bfseries}
\subsubsectionfont{\normalfont\normalsize\scshape}

\makeatletter
\newcommand{\quotefont}{\fontfamily{qtm}\selectfont\upshape}
\let\oldquote\quote
\let\endoldquote\endquote
\renewenvironment{quote}{\oldquote\small\singlespacing\quotefont}{\endoldquote}
\let\oldquotation\quotation
\let\endoldquotation\endquotation
\renewenvironment{quotation}{\oldquotation\small\singlespacing\quotefont}{\endoldquotation}
\makeatother

\numberwithin{equation}{section}

\begin{document}

\sloppy

\title{Is quantum mechanics merely a theory for us?}

\author[1]{\bf Peter W. Evans\thanks{email: \href{mailto:p.evans@uq.edu.au}{p.evans@uq.edu.au}}}

\affil[1]{\small{{\it School of Historical and Philosophical Inquiry}, University of Queensland}}

%\author{Peter W. Evans}

%\institute{P. W. Evans \at
%              School of Historical and Philosophical Inquiry \\
%              University of Queensland \\
%              St. Lucia, QLD 4072, Australia \\
%              Tel.: +61-7-336-52162\\
%              \email{p.evans@uq.edu.au}\\
%              ORCID: 0000-0003-0214-4748
%}

\date{}

\maketitle

\begin{abstract}
  This paper develops an agent-centric account of measurement that treats the preferred-basis problem is fundamentally perspectival. On this view, the system--apparatus--environment decomposition and the observables that are apt to become classically robust are determined by the physical constitution and epistemic constraints of an embodied class of agents. Decoherence then stabilises those agent-specified observables, yielding facts that are \emph{stable for us} without positing an absolute, observer-independent basis. On this picture, `measurements' are public not because they are metaphysically privileged, but because agents like us share the relevant sensorimotor and operational structure. I motivate this account through a discussion of two recent no-go results for relational quantum mechanics (RQM) \citep{Brukner21,Pienaar2021}, and a subsequent response \citep{DiBiagio2022}: my aim is not to defend RQM \emph{per se}, but to refine the relational insight with a principled account of basis selection rooted in embodiment. I provide a phenomenological gloss, drawing on body-schema considerations, to argue that quantum mechanics is best understood as an idiosyncratically human description of interactions with the physical world---a structurally constrained, agent-indexed framework within which classicality emerges.
  %\keywords{Relational quantum mechanics \and Decoherence \and Preferred-basis problem \and Agent-centric quantum mechanics}
\end{abstract}

 \tableofcontents

\section{Introduction}

A central challenge for quantum foundations is to explain how a specific set of definite outcomes (a preferred basis of eigenstates) emerges from quantum superposition. Relational quantum mechanics (RQM) answers this challenge by denying that quantum properties are absolute, instead claiming they exist only \emph{relative} to interactions between systems. Any interaction between an `observer' and an `observed' counts as a measurement, and the outcome is a fact only for those systems. There is no God's-eye perspective; reality is a network of relative quantum facts. This immediately raises a question: if facts are defined only relative to each observer, how can multiple observers ever agree on what happened in a given quantum process? What ensures that relational perspectives can be made consistent?

Two recent no-go theorems from \citet{Brukner21} and \citet{Pienaar2021} sharpen this challenge. In short, without a built-in mechanism to align different observer perspectives, a strictly relational framework seems to invite mutually contradictory descriptions of reality. At the heart of these challenges lies the preferred-basis problem: what determines the particular basis in which quantum superpositions resolve into concrete outcomes? In conventional quantum theory, environment-induced decoherence provides a partial answer by dynamically selecting a stable set of `pointer' states: the environment effectively measures certain observables of the system, suppressing interference and thus defining a set of states that behave classically. But in the relational context of RQM there is no single objective environment or fixed frame that \emph{universally} picks out a basis. Decoherence occurs relative to interactions; it does not, by itself, fix a common basis across observers.

The narrative begins in \S\ref{sec:nogo} with the no-go results and a recent response by Di Biagio and Rovelli as a motivation for considering the preferred-basis problem in this new light. However, the main goal of this paper is not to defend RQM \emph{per se} in the face of these challenges, but to develop an \emph{agent-centric} account on which the preferred-basis problem is fundamentally perspectival. In \S\ref{sec:agcentprefbas} I propose that the system--apparatus--environment decomposition and the observables that are apt to become classically robust are determined by the physical constitution and epistemic constraints of an embodied class of agents. Decoherence then stabilises those agent-specified observables, yielding facts that are \emph{stable for us} without positing an absolute, observer-independent basis. This is not to say that agents \emph{cause} decoherence, but that decoherence becomes explanatory only relative to the decomposition afforded by the agent perspective. The agent perspective defines the set of dynamical variables and the kinds of system--environment splits that characterise that agent's interaction with physical systems \citep{Evans20,Kearney2020}. This agent-dependence offers a natural route to consistency between different observers: if two agents in differing relative contexts compare their observations, the act of comparison is itself a measurement interaction that creates a joint context---for instance, a shared communication channel or environment---that makes a coherent comparison possible. Thus, on this picture, `measurements' are public not because they are metaphysically privileged, but because agents like us share the relevant sensorimotor and operational structure.

As a result, in \S\ref{sec:agcentqm} I argue that quantum mechanics is best understood as an idiosyncratically human description of interactions with the physical world---a structurally constrained, agent-indexed framework within which classicality emerges. Consequently, I contend that it is a misplaced presumption that the interaction between two arbitrary physical systems is appropriately described by quantum mechanics independently of any detection and measurement by (an instrument of) a human agent. \S\ref{sec:embodphenqm} builds on this to provide a phenomenological gloss, where I am emboldened by recent phenomenological approaches to quantum mechanics that emphasise the agent's embodied role in defining events \citep{Berghofer2023,French2024}. In short, my proposal rests on a phenomenological insight that complicates any neat distinction between ontology and epistemology. Drawing on Merleau-Ponty, the structure of the physical world is not a fixed backdrop passively accessed by disembodied observers; it \emph{emerges} through the relational dynamics of agent--world coupling. In this context, the agent's perspective is not merely a standpoint from which pre-existing facts are measured; it is the condition under which certain features of the world become actualised as facts. The classical structures we attribute to quantum systems---including the delineation between system and environment, and the selection of a preferred basis---are co-constituted through the embodied and operational constraints of the agent. This view reframes decoherence not as a purely physical process awaiting external description, but as an enacted phenomenon: it is the formal signature of stability emerging from within a specific interactive context. Adopting this stance relinquishes the `view from nowhere' and treats quantum theory as a structured description of agent--world interaction -- one that resists the clean separation of metaphysical structure from epistemic access.\footnote{For phenomenological readings that connect quantum theory to situated embodiment---and, in particular, to Merleau-Ponty---see Bitbol's discussion of QBism and endo-ontology (articulated from within embodied agency) \citep{Bitbol2020} and the eco-phenomenological development (that stability and meaning are achievements of agent--environment coupling) with de la Tremblaye \citep{BitbolDeLaTremblaye2023}. My proposal is complementary: it retains the phenomenological emphasis on situation while specifying the physical pathway (via agent-conditioned decompositions and decoherence) through which classical stability is secured for a class of embodied agents.} Let us begin with the no-go theorems.

\section{Relational quantum mechanics and the no-go theorems}
\label{sec:nogo}

Posted online mere days apart in 2021, \citet{Brukner21} and \citet{Pienaar2021} independently formulated no-go theorems challenging the ability of RQM's to maintain consistency across relational frames of reference. Though some of the details differ, both conclude that RQM lacks structural resources to ensure consistency in how quantum states are assigned and updated between observers. Let us begin here with a brief recap of the the main claims of RQM.

According to RQM, quantum properties (such as the value of an observable) are not intrinsic attributes of an isolated system -- they manifest only upon interaction with another system. Any such interaction is considered a measurement, and so the terms `observer' and `observed' can be applied ubiquitously across quantum systems. Reality on this view is constituted by interactions---quantum events---in which systems affect one another and acquire definite relational properties, forming a `sparse' ontology of the world \citep{RovelliSEP}. Thus reality is, in a sense, observer-dependent: a measurement outcome is a fact only for the systems that interact.

RQM does not formally diverge from standard quantum mechanics, but it does explicitly reject wavefunction collapse. An interaction implies loss of coherence for the observer interacting with the system, but not necessarily for other potential observers. Accordingly, the wavefunction is interpreted epistemically rather than as an intrinsic description of reality. Measurement is simply an interaction that produces correlations between systems, where decoherence and information exchange between systems define the outcomes of the interaction within a given perspective. A key implication of this is that different observers have access to legitimately different information about their immediate environment until they interact and exchange it -- there is no omniscient standpoint from which all facts combine into a single absolute truth about reality. The core of this ``radical perspectival antifoundationism'`' is that \citep{RovelliSEP}:
\begin{quote}
  The probability distribution for (future) values of variables relative to a system $S$ depend on (past) values of variables relative to the same system $S$, but not on (past) values of variables relative to another system $S^{\prime}$.
\end{quote}
This is the point at which the no-go theorems place pressure. It is worth noting, however, that there is a growing discussion of how RQM might defend inter-observer consistency \citep{Adlam2022,AdlamRovelli2023,Cavalcanti2023}. Whilst I do not address these wider discussions explicitly, the account developed below might be seen as a contribution to this debate. Let us now turn to the first no-go theorems.

\subsection{Brukner's no-go theorem}

Since in RQM any interaction can count as a `measurement', even a single qubit could in principle play the role of observer. \citet{Brukner21} poses the question: does every such `observer' thereby have `knowledge' of the observed system? He derives a no-go result: under reasonable assumptions, a quantum observer like a qubit cannot consistently possess well-defined knowledge of another system without contradiction. Brukner's key step is to identify in RQM an ``ambiguity in the choice of the basis with respect to which the relative states are to be defined'' \citep[p.~1]{Brukner21}. Consider an `observer' $O$ and system $S$ in a joint state $\ket{\psi}_{SO}$. Brukner's theorem turns on two assumptions \citep[p.~2]{Brukner21}. First, (DefRS): for \emph{any} convex decomposition
\[
\ket{\psi}_{SO}=\sum_i c_i \ket{x_i}_S\ket{X_i}_O,\qquad \sum_i |c_i|^2=1,
\]
the occurrence of $\ket{X_i}_O$ (a ``state of knowledge'') implies that $S$ is in the \emph{definite relative state} $\ket{x_i}_S$. Second, (DisRS): distinct relative states of $S$ correlate with orthogonal observer ``knowledge'' states, so if $\ket{x}_S\neq \ket{x'}_S$ then $\langle X|X'\rangle=0$. Together these encode that $O$ can `know' the state of $S$ (relative to some basis), with each distinct $\ket{x_i}_S$ recorded in a distinct, orthogonal memory state $\ket{X_i}_O$.

The contradiction appears in the simplest nontrivial case. Let $S$ be spin-$\tfrac{1}{2}$ and $O$ two-dimensional. After interaction, $\ket{\psi}_{SO}$ may be written in the $\{\ket{\uparrow},\ket{\downarrow}\}$ basis so that, by DefRS, $O$'s states encode `$S$ up' ($\ket{O_{\uparrow}}$) and `$S$ down' ($\ket{O_{\downarrow}}$), which DisRS requires to be orthogonal. The same $\ket{\psi}_{SO}$ can also be expanded in the incompatible $\{\ket{\leftarrow},\ket{\rightarrow}\}$ basis for $S$, yielding corresponding ``knowledge'' states $\ket{O_{\leftarrow}}$ and $\ket{O_{\rightarrow}}$ that DisRS again demands are orthogonal. But $\ket{O_{\leftarrow}}$ and $\ket{O_{\rightarrow}}$ are convex combinations of $\ket{O_{\uparrow}}$ and $\ket{O_{\downarrow}}$ and are, in general, \emph{not} orthogonal to them. Hence $O$ cannot simultaneously satisfy DisRS for two incompatible choices of basis on $S$. Once $O$ is correlated with $S$ in one basis, $O$ cannot also be described as having definite knowledge in another without contradiction.

At least one of DefRS or DisRS must therefore fail. Mere correlation does not amount to basis-independent `knowledge' of the system. As Brukner observes, this is a version of the preferred-basis problem: it is ``consistent with the view that the states of knowledge of the observer are defined\dots in only one, preferred, basis'' \citep[p.~2]{Brukner21}. RQM's permissive stance---any interaction is a measurement, with no basis specified---leads directly to this ambiguity. Thus not all quantum systems can serve as observers \emph{in this strong sense}: only systems with well-defined, distinguishable record states---typically macroscopic measurement devices or agents---can function as observers possessing definite knowledge about other systems. A qubit entangled with a qubit may instantiate a relation, but not a singular classical outcome we could call `the qubit's observation'.

As Brukner notes, this contrasts with neo-Copenhagen and QBist approaches, where `measurement' and `observer' are more narrowly framed and a preferred basis  determined by the experimental context or the beliefs of the agent \citep[p.~3]{Brukner21}. While decoherence can dynamically \emph{select} stable pointer observables, the absence of an intrinsic preferred basis in RQM means decoherence alone does not yield a basis-independent notion of `knowledge'. Brukner's theorem thus shows that such relational state assignments cannot be reconciled in a way that preserves the consistency of quantum mechanics across different observational frames.

\subsection{Pienaar's no-go theorem}

Pienaar's challenge parallels Brukner's by employing a variant of Wigner's friend: at `Event 1', the friend $F$ measures a system $S$ and records an outcome; at later `Event 2', the superobserver $W$ measures the joint $FS$ system. Pienaar claims RQM is committed to ``shared facts'' between $W$ and $F$ (\textbf{RQM:6}): any outcome $W$ obtains on $FS$ should be consistent with $F$'s outcome on $S$. He then articulates \textbf{P1}: $W$ is free to measure in \emph{any} basis at Event~2; and \textbf{P2}: any basis-specific state update on $FS$ relative to $W$ at Event~2 indicates that $F$ obtained an outcome in the \emph{same} basis on $S$ at Event~1.

To generate a tension, Pienaar chooses the initial measurement at Event~1 to be in a basis \emph{different} from $W$'s basis at Event~2. Given \textbf{P1}, $W$'s basis cannot be constrained by $F$'s earlier choice. Then \textbf{P2} effectively makes the Event~1 relation depend on the Event~2 choice: the basis (and hence content) of the $FS$ correlation appears fixed by $W$'s later decision. But this violates the core RQM commitment \textbf{RQM:3}: relations are intrinsic to the systems involved and ``independent of anything that happens outside these systems' perspectives'' \citep[p.~5]{Pienaar2021}. On Pienaar`'s reading, the $FS$ relation at Event~1 would be retroactively determined by $W$'s later measurement.

Pienaar concludes with a trilemma: \textbf{P1}, \textbf{P2}, and \textbf{RQM:3} cannot all be true \citep[p.~23]{Pienaar2021}. Each horn goes against the spirit of RQM:
\begin{enumerate}
  \item \label{enum:i} rejecting \textbf{RQM:3} denies that relations are intrinsic to an interacting pair;
  \item \label{enum:ii} rejecting \textbf{P1} suggests hidden-variables-like constraints on $W$'s measurement freedom;
  \item \label{enum:iii} rejecting \textbf{P2} implies that not all physical correlations qualify as measurements.
\end{enumerate}
Horn \ref{enum:iii} converges with Brukner's result: not all quantum systems are observers in the strong sense, and so not all interactions constitute measurements. The preferred-basis pressure resurfaces: if a fixed basis objectively emerged, this would underpin \textbf{P2}; but RQM's observer-relativity makes a unique basis difficult to sustain. Without principled basis-fixing, equally valid but incompatible descriptions may be unreconcilable between different observers, with no rule selecting a final basis for an integrated account. Thus Pienaar complements Brukner in showing decoherence alone is insufficient: it explains \emph{stability} within a context but does not, by itself, dictate which basis is fundamental \emph{across} perspectives. If observers deploy different bases, their accounts need not be transitively alignable, which undermines the coherence of RQM.

\subsection{A response in defence of RQM}
\label{subsec:rqmresponse}

\citet{DiBiagio2021,DiBiagio2022} contend that the no-go theorems sharpen, rather than undermine, RQM's distinctiveness by foregrounding its radical relationalism: facts, not quantum states, are primary. Their key move is to distinguish relative facts from stable facts---the latter being those subsets of relative facts that are jointly relative to (and retrievable by) the \emph{same} third system. On their view, stability is supplied by decoherence.

Against Brukner, they argue that the tension originates in projecting classical `knowledge' onto bare quantum correlations. In RQM, entanglement between systems does not automatically constitute knowledge in any agent-centric sense. Either one treats `knowledge' as mere correlation (in which case orthogonality need not hold), or one insists on orthogonality and thereby restricts `knowledge' to decohered, macroscopic records. In neither case does RQM ascribe epistemic states to generic quantum systems, so the alleged inconsistency does not arise.

Against Pienaar, they accept \textbf{P1} and \textbf{RQM:3} but reject \textbf{P2}. On their reading, Pienaar's inference---that a post-selected state for $W$ on $FS$ indicates that $F$ measured $S$ in the same basis---mischaracterises the role of states in RQM. Quantum states are not ontic descriptions; they are devices for calculating probabilities of events relative to a system. The correlation observed when $W$ measures $FS$ need not imply anything about $F$'s earlier basis; it may simply reflect how $W$ chose to couple to $FS$. More generally, they rearticulate RQM's commitments to clarify when correlations count as facts without presupposing `measurement', and they stress that shared facts arise only through explicit interactions that permit cross-perspective comparison. This blocks third-party inferences about what occurred between two systems prior to any such interaction.

I do not here adjudicate whether this constitutes a coherent defence of RQM. Rather, I leverage the above concerns to motivate my account that the preferred-basis problem is inherently perspectival: basis selection is constrained by the agent's physical constitution and epistemic limitations, and different agents have access to different interactions, which fix their effective preferred bases. This reframes decoherence and measurement without positing an ontologically fundamental preferred basis, and it aligns with Brukner's insight that outcomes must be stabilised in some basis---now understood as agent-dependent rather than system-intrinsic.

Taken together, these results are best read as a diagnosis of what a purely relational picture leaves open. The issue is not whether facts are absolute, but which interactions yield records that persist and can be compared without inviting contradiction. The remainder of the paper develops a positive thesis: for agents like us, the relevant system--apparatus--environment decomposition and the observables that behave classically are fixed by how we are built and how we couple to the world; decoherence then stabilises exactly those features. Section~\ref{sec:agcentprefbas} sets out this agent-centric account of basis selection. Section~\ref{sec:agcentqm} argues that, as a result of these considerations, quantum mechanics is best understood as an idiosyncratically human description of interactions with the physical world. Section~\ref{sec:embodphenqm} then offers a phenomenological gloss, according to which the structure of the physical world is not a fixed backdrop passively accessed by disembodied observers; it \emph{emerges} through the relational dynamics of agent--world interactions.

\section{An agent-centric account of measurement}
\label{sec:agcentprefbas}

In this section I want to bring together a small cluster of different ideas to propose an agent-centric approach to the preferred-basis problem. The first part of this involves rehearsing an argument I make in \citep{Evans20} that an agent's objectification of structure in the world is contingent upon their physical constitution and epistemic limitations. This feeds into the second part that involves drawing out an agent-centric approach to decoherence, based on the lesson from the first part. The approach aligns with phenomenological readings of quantum theory (for instance, from \citet{Bitbol2020}) that I explore in greater depth in \S\ref{sec:embodphenqm} while adding a concrete mechanism---agent-conditioned decompositions plus decoherence---to explain why classical stability is expected for agents like us. Let us begin with a very brief review of the orthodox narrative for how decoherence contributes to understanding the preferred-basis problem.\footnote{In what follows, I adopt the vocabulary of `measurement', `observer', and `fact' in a way that is broadly consistent with the usage in RQM -- where any physical interaction can constitute a measurement, and any system can serve as an observer. However, the agent-centric framework I develop here departs from this permissiveness. In this view `observers' are not just any systems, but embodied agents with specific structural and operational constraints; `measurements' are not every interaction, but those capable of stabilising decohered outcomes relative to such agents; and `facts' are not simply relational events, but emergent structures that become accessible through the interactional affordances of the agent. These distinctions become especially salient when we consider decoherence and system--environment delineation.}

\subsection{Decoherence and the preferred-basis problem}

Quantum decoherence is a mechanism that explains the emergence of classicality from quantum systems due to their interaction with the environment: interactions between a system and its environment lead to the suppression of the interference terms in the quantum state, which effectively singles out a preferred basis in which the system appears classical. Here is \citet{Schlosshauer05} describing this suppression effect:
\begin{quote}
  In the majority of the cases accessible to our experience\ldots interaction with the environment is so dominant as to preclude the observation of the `pure' quantum world, imposing effective superselection rules\ldots onto the space of observable states that lead to states corresponding to the `classical' properties of our experience. Interference between such states gets locally suppressed and is thus claimed to become inaccessible to the observer.
\end{quote}

This process of environment-induced superselection, or `einselection', determines the preferred basis by dynamically selecting the quantum states (the `pointer states') that remain stable under environmental interactions \citep{Zurek1982}. The crucial point is that the preferred basis is not imposed externally but emerges from the dynamical properties of the interaction Hamiltonian, which defines how the system interacts with its surroundings. The stronger the coupling with the environment, the faster decoherence suppresses off-diagonal elements in the system's density matrix, leading to classical-like behaviour. The pointer states that arise from this diagonalisation can be identified as approximate eigenstates of certain system observables that commute with the interaction Hamiltonian, and so define a preferred basis since they remain robust under decoherence. This aligns with the notion that the environment is effectively measuring those observables.

The mechanism of decoherence has been supplemented in recent years by the framework of quantum Darwinism \citep{Zurek2009}, which attempts to explain how classical reality emerges objectively in the decoherence process. According to this view, the environment not only decoheres the system, but it also selectively proliferates information about the system's pointer states throughout many parts of the environment. The environment acts as a communication channel, or `witness', encoding redundant information about the system's state. Only `robust' states, in which the system--apparatus correlations are stable under environmental interactions (such as pointer states), and that can be efficiently copied by the environment, become accessible to observers. Through what Zurek describes as the `predictability sieve', the most stable states will also be the most predictable states since they maintain maximal redundancy in their environmental imprint and so minimise information loss to the environment, and so any less robust states `die out' (hence the analogy with Darwinism). It is the fact that this selection is driven entirely by the interaction between the system and environment that renders this an objective physical process.

We will come back to this issue of objectivity in a moment. Let us now consider how the agent objectification of structure in the world could be contingent upon the physical constitution and epistemic limitations of an agent.

\subsection{Agent-centric objectification}
\label{subsec:agcentobj}

In \citep{Evans20}, I promote the notion that quantum mechanics displays all the hallmarks of agent-centric objectification by analysing the parallel case of the objectification of causal relations in a view known as `causal perspectivalism' \citep{Price07}. The key to the argument in the case of causal relations in the interventionist framework \citep{Woodward03} is as follows. Upon establishing through interventions on some physical system that a functional relation exists between some pair of variables we take to characterise the system, we call this functional relation a causal relation when it meets a specific set of interventionist conditions \citep[p.98]{Woodward03}. These conditions include that the functional relation is invariant and stable relative to a range of interventions and background conditions, and this is connected in the interventionist account to the idea that the level of detail or generality of the variables that we take to characterise these functional relations---the `level of grain'---in a sense is already stipulated. Thus, whether we have uncovered a causal relation via our intervention on some system depends on our particular coarse-graining of the system, and we undertake this coarse-graining when we model physical systems \emph{just so} dynamical variables with the right sort of functional interrelationships can be objectified for our practical purposes \citep[p.10]{Evans20}.

When these considerations are paired up with Price's claim that the distinction between cause and effect can be reduced to the agent perspective, through ``the strong temporal bias of our epistemic access to our environment'' \citep[p.280]{Price07}, we see precisely how an agent's causal ascriptions to the world are contingent upon their physical constitution---how they objectify variables in the process of coarse-graining---and epistemic limitations---how they are oriented in time. More than this, according to \citet{Ismael16} agents pragmatically partition the world into those parts that are exploitable for interventions and those that are not---that is, into cause and effect---based on the agent's idiosyncratic epistemic constraints and limitations concerning the `frame-invariant' modal structure of the world. In doing so, the agent makes a model-based pragmatic separation of the world into system and environment as a function of the agent's ability to objectify causal variables and relations.

My argument in \citep{Evans20} goes further than this to claim that such perspectival objectification can be applied in the quantum context also. The key to this argument stems from Giere's claim that scientific instruments are perspectival: instruments are sensitive to a particular kind of input and are blind to everything else \citep[p.14.]{Giere06}. In so far as the constrained idiosyncratic capabilities of agents to interact with and model the world can be thought of as `instruments', then the physical capabilities of agents define an agent perspective in the same sense. As I put it in \citep[p.16]{Evans20}:
\begin{quote}
  the interaction between an instrument and a system characterises a kind of `capability' for the instrument by constraining the subset of variables to which the measured system can respond. But we design and engineer scientific instruments to mesh with our perceptual capabilities, thus we manufacture overlap between the `capabilities' of our scientific instruments and our idiosyncratic sensory contingencies. So in the quantum context\ldots the experimental arrangement is specifically devised by the agent to bring about the phenomenon in accordance with the idiosyncratic experiential faculties of that agent.
\end{quote}

As a result of these considerations, we can see how `instruments', including an agent's own instruments that are part of its physical constitution, as well as any further engineered instruments that concatenate with the agent's own, inherently define a set of variables to which the instruments are sensitive, delineate the world into system and environment based upon this set of variables, and disambiguate the functional relations that can be objectified at the level of grain commensurate with that sensitivity. So there is a plausible sense here in which the specific operation of an instrument in the quantum context can supply the system--environment split and stable basis required for the decoherence mechanism to proceed. The more revolutionary (and admittedly speculative) implication of this, however, is that in so far as we, as human agents, and our interactions with the world are defined by our own idiosyncratic physical constitution, the variables to which we are sensitive, the typical ways we delineate system and environment, and the kinds of levels of grain that are most useful for us are all based on our characteristic ability to intervene on the world, and so these factors in effect render the theory of quantum mechanics a peculiarly human theory of agent--world interactions.

Although I do not make the following connection in \citep{Evans20}, preferring instead to invoke Bohr's philosophy of quantum theory, this agent-centric approach provokes phenomenological insights---particularly those of \citet{MerleauPonty1945}---in which the traditional divide between epistemology and metaphysics is dissolved. Perception is not a passive reception of facts about the world, nor is it merely an internal representation; it is a mode of \emph{being-with} the world in which objectivity emerges only through embodied interaction. In this context, the agent's epistemic access is not merely perspectival knowledge of an independent ontology; it \emph{constitutes} the relevant ontology. I will say more about this in \S\ref{sec:embodphenqm}. For now, let us connect this more specifically to decoherence and the preferred-basis problem.

\subsection{Agent-centric decoherence}

\citet{Kearney2020} argues for the agent dependence of the emergence of pointer states in the decoherence process. According to decoherence, when a quantum system $S$ interacts with an apparatus $A$ and environment $E$, their joint state evolves into an entangled superposition. Tracing out the environment yields a reduced density matrix for $SA$ that is approximately diagonal in some pointer basis. Despite the fact that this reduced state is an improper mixture, Kearney argues that the reduced state can be treated as if it were a proper mixture on account of the fact that any interference between outcome states is locked into environmental degrees of freedom and would require a highly improbable `recoherence' measurement on the entire $SAE$ composite to detect. As such, for all practical purposes an agent restricted to local measurements on $SA$ will obtain statistics indistinguishable from those of a true classical mixture, enabling the diagonalised reduced state to be treated as an ignorance-interpretable distribution of outcomes.

A key part of this account is the need for an agent to delineate the system from its environment, which in turn induces a pointer basis within which decoherence occurs; quantum mechanics itself does not dictate a unique way to factor the world into subsystems. In the application of the decoherence mechanism, agents (through their apparatus) choose which degrees of freedom constitute the system of interest and so also what constitutes the environment. Different choices of this delineation lead to different descriptions of which observables decohere. As \citet[p.23]{Kearney2020} notes, this holds for the accounts of stability and predictability in the quantum Darwinism program. As such, a preferred basis arising through the mechanisms of quantum Darwinism can only be derived once a decomposition of the whole $SAE$ Hilbert space is chosen: that is, the partition of the Hilbert space is chosen relative to what an agent considers relevant or can access. This pushes the question about a preferred basis back to a question about a preferred decomposition into subsystems. This decomposition is not dictated by quantum mechanics: at best ``the distinction between system and environment seems arbitrary'' \citep[p.3635]{Tanona2013}, and at worst ``it depends on the set of resources effectively available to access and control the degrees of freedom of S'' \citep[p.4]{Zanardi2001}. These quotes suggest not merely that the decomposition is flexible, but that it is operationally constrained -- in this context by the nature of the instrumentation employed. However, as I have argued above, such instruments are designed precisely for their use by embodied agents, as a kind of proxy for the physical constitution of the agent, and so the decomposition is essentially functionally determined by the physical and cognitive structures that mediate the agent interaction with the world.

In this way, the issue of a preferred basis comes hand in hand with the delineation of system and environment. Decoherence naturally picks out a basis (the pointer states) in which the reduced density matrix is diagonal (or nearly so), corresponding to robust, stable states that do not suffer interference due to environmental monitoring. But we might ask, why do we see pointer states and not some other basis? The quantum Darwinism program would proffer the pragmatic reply that only pointer states survive decoherence long enough to be observed, and creatures like us need stable information to survive, so we evolved to perceive pointer states: very roughly, our sensory apparatus co-evolved with our intuitive recognition of the `classical' world, with quantum states that are predictable and robust forming the basis of perceived reality. But given what we have said above, all these arguments presuppose that the world has already been delineated into system and environment.

It is at this point that the considerations in \S\ref{subsec:agcentobj} become significant. As we have just seen with respect to the objectification of causal relations (and what are our physical theories if not causal models of the world?), agent attributions of causal structure to the world are highly contingent on the physical capabilities and epistemic limitations of the agent. In particular, where an agent draws the distinction between the system of interest and the surrounding environment depends upon to which variables the agent's idiosyncratic `instruments' are sensitive, and which level of coarse-graining objectifies exploitable functional relations between those variables. Since these factors are determined by the physical constitution of the agent, then this feature of the agent also determines the nature of any system--environment split, and so the emergence of a pointer basis. In effect, depending upon what we `ask' of a system (according to the instrumentation we apply, either natural or artefactual), we objectify both different system--environment delineations and different sets of variables characterising the system, and so define different possible bases.\footnote{In addition to this, one might argue that the redundancy required for predictability and stability itself presupposes an agent who is capable of accessing that redundant information and distinguishing it from noise. This provides an extra dimension to the claim that the emergence of the classical world according to quantum Darwinism is not absolute but rather relative to agents embedded within a structured environment.}

On this view, decoherence is not an observer-independent process that agents happen to describe in one basis or another. Rather, decoherence is the formal realisation of a more fundamental enactive relation -- the stabilisation of facts through embodied interaction with the world. The agent's physical capacities \emph{are} the structure through which a particular system--environment split becomes actualised, and hence through which classicality emerges. There is no `view from nowhere': decoherence is what interaction looks like \emph{from within} an embodied relation. As such, an agent is not free to choose any arbitrary delineation between system and environment, as this delineation is a function of a feature of the agent that is outside of the agent's control (modulo the agent's ability to develop instruments that extend their physical capabilities -- but even then these instruments would be constrained to concatenate with the constitution of the agent). On the flipside, the physical system itself conveys, in the words of \citet{Weinberger2024}, a `worldly infrastructure' that contributes affordances that shape the agent's interaction with it.

Moreover, an equivalence class of agents would share their physical and epistemic constraints: they would all be made of the same physical stuff operating at the same scales using the same tools.\footnote{With co-authors, we elaborate on this issues in \citep{EMS2021}} By `equivalence class' here, I refer to agents whose physical and cognitive architecture---embodiment, scale, sensory modalities, and extended instruments---are sufficiently similar to produce comparable interactions with the world. Humans constitute such a class; whether artificial systems or non-terrestrial lifeforms would belong depends on shared operational constraints. As a result of this, an equivalence class of agents would share the same natural preferences for delineations of target systems into system and environment, and so given the account above would share the same emergent classical reality. In any given experimental scenario, a shared preferred basis would effectively emerge between agents of the same equivalence class as a result of environmental decoherence via quantum Darwinism.

If this story is right, the upshot of it all is that the selection of a preferred basis is not an absolute physical fact but a contextual feature of how agents structure and interpret quantum interactions -- quantum theory becomes an idiosyncratically human description of our interactions with the physical world.

\section{Agent-centric quantum mechanics}
\label{sec:agcentqm}

Let us return to the two no-go theorems of \S\ref{sec:nogo}. The no-go theorems attempt to make trouble for RQM with respect to the ability of the account to define under what circumstances a measurement actually occurs, particularly so when it comes to disparate observers who may relate to the target system in completely different ways. Despite the fact that the considerations of the previous section appear to go against the spirit of RQM, in that they explicitly bring agents into the description of quantum phenomena, I wish to use these considerations to proffer here a response to the challenges that Brukner and Pienaar raise for RQM.

Both no-go theorems put pressure on the idea from RQM that all interactions between physical systems constitute valid quantum measurements, such that not all quantum systems can serve as observers. The account I have espoused here explicitly provides a context-dependent mechanism for basis selection: the physical constitution and epistemic limitations of the agent underpin any objectification of the variables of interest in the target system and any system--environment split. By grounding preferred-basis selection in the interaction constraints of the agent, this approach ensures that state assignments are not arbitrary but determined by pragmatic and operational considerations. As a result, whether or not an interaction counts as a measurement ends up defined relative to these agent-centric considerations.

There is still an outstanding issue to be solved here, which is of course just the measurement problem itself, concerning which interactions count as agent-based and which not. As such, I do not propose to have addressed the measurement problem by making this move, only to have rearranged the moving parts of the problem to look a little differently. I will say a few words below about what I take the philosophical implications of this move to be.

We mentioned in \S\ref{subsec:rqmresponse} above a potential response to the no-go theorems on behalf of RQM, in which a distinction was drawn between the complete set of relative facts that emerge across the whole network of relations between interacting physical systems, and the subset of stable facts within this complete set, which emerge through the process of decoherence as shared facts amongst some class of observers who all relate to the same target system \citep{DiBiagio2021,DiBiagio2022}. In this section, I wish to consolidate the ideas presented thus far to provide a plausible account for why we should think of (the accessible subset of) these shared stable facts according to RQM as comprising a peculiarly human way of seeing the quantum world. The account I present here leverages the considerations of the previous section to argue that the broad contours of the quantum world we observe are underpinned by our own physical constitution and epistemic limitations, our own agent perspective, which together determine the set of functionally meaningful variables available for measurement and interaction. We pick this thread up again in \S\ref{sec:embodphenqm}. Before that, let us say a few more words about the no-go theorems.\footnote{This proposal of an `agent perspective' has some obvious similarities to Healey's concept of an `agent situation', in which the ``ascription of a quantum state to a system relates that system to a physically characterised situation that may (but need not) be occupied by a physically situated agent'' \citep[p.752]{Healey2012}. Healey argues that an agent's observational and inferential capacities frame their interaction with quantum systems, such that what counts as a `measurement' is contingent on their situation. Healey's account provides a nice explanation for why we might ascribe a quantum state to the interaction between otherwise undisturbed arbitrary physical systems, such as those considered above. This is because we need only imagine a physically situated agent accessing the relational information in the interaction to ascribe to it a quantum state.}

\subsection{Is quantum mechanics merely a theory for us?}

The window into this discussion I would like to employ is by asking the rather facetious question that is the title of this paper. When two arbitrary physical systems interact---one serving as the `observer' and the other as the `observed' system---what determines the basis in which their interaction occurs? Moreover, does the `observer' system model its interaction with the `observed' system using quantum mechanics at all? Of course, `modelling' is an anthropocentric concept, and arbitrary physical systems are not in the business of modelling anything in the sense that we do. But does the interaction between these arbitrary systems obey the constraints of quantum mechanics? And how might we know this? Let us explore this in greater detail.

According to Di Biagio and Rovelli, any pair of physical systems that interacts defines a relative fact, where one system provides the `context' for ``the fact that a certain variable [characterising the other system] has a value in that context'' \citep[p.3]{DiBiagio2021}. According to standard quantum mechanics, when two systems interact their joint state typically evolves into an entangled pure state in the combined Hilbert space. Either subsystem is then described by a reduced density operator, obtained by tracing out the degrees of freedom of the other. In general, this reduced state is a mixed state. Significantly, any given mixed state admits infinitely many convex decompositions into pure states. Likewise, a pure entangled state can be expressed as a sum over product states in infinitely many ways. Whilst the convex decompositions of a mixed state are not determined by any particular basis, the different product-state decompositions of a pure entangled state do depend on the choice of basis in the composite Hilbert space. In the absence of additional physical structure---such as a measurement context or an environment---there is no uniquely preferred basis for such decompositions.

Since in RQM one system provides the context for the other, this suggests that a choice of basis---determined by the interaction and the relational structure---becomes manifest in the correlation established between them. If one physical system has enough degrees of freedom to function as an effective environment, such as a measurement device, the interaction may begin to decohere the relative state of the other system into a set of stable states in some preferred basis.\footnote{I do not claim that decoherence is necessary for relational facts to exist in the RQM sense. Rather, I claim that decoherence---understood through the lens of agent-centric constraints---is the mechanism by which such facts become stabilised and accessible across perspectives, thus enabling consistency and classicality within an equivalence class of agents.} But if the interaction remains coherent and undisturbed by external degrees of freedom, no stable basis emerges, and the facts of the interaction remain relative and accessible only to the two systems involved.

Now imagine probing the system in an attempt to access the relative facts established during this previously undisturbed interaction. If the emergence of a preferred basis during measurement depends in part on the structure of the measurement interaction---including the instruments we employ and variables we are physically capable of objectifying, as I argued in the previous section---then any subsequent measurement risks transforming the delicate relational correlations into a new set of stable facts. These stable facts reflect not just the prior relational structure, but also the agent-centric constraints imposed by the measurement process. In this sense, the act of measurement does not simply reveal pre-existing relative facts but actively reshapes the informational landscape, such that only those features which couple to our particular modes of interaction become stably accessible. As a result, our epistemic access to the original relational facts is limited: we can only retrieve what survives the transition into a form compatible with our own embodied measurement context.

Of course, this is not to rule out the possibility of, perhaps, some form of quantum tomography to probe and recover relational data without destroying it, and so maybe it is too hasty to say that relational facts are fundamentally unknowable unless stabilised in a shared environment. But there is at least some sense here where the explanatory power of RQM comes under a small amount of pressure: we potentially lose epistemic authority to say precisely what happens where we have no operational access. Quantum systems may not model their environments using quantum mechanics.

I do not want to push this speculative point too hard. I imagine a defender of RQM simply denying the agent-centricity of the system--environment split, for instance. Indeed, one might object that this view conflates epistemic perspective with ontological emergence, thereby undermining any robust account of quantum structure. But making such a suggestion contains a key presupposition of a clean separation between knowing and being -- a separation which, in phenomenological terms, is itself a product of pure abstraction. On the view defended here, the structure of quantum phenomena is not constructed by agents, nor discovered by them, but co-emerges through the relational structure of embodied interaction. Thus, one way to respond is that, rather than undermining relationalism, the above agent-relative stance provides a novel set of structural resources needed to ensure consistency in state assignments and updates between observers.

Consequently, whilst the agent-centric account I develop here may seem to depart from the egalitarian spirit of RQM---in which all physical systems can serve as observers---it need not be taken as a repudiation of the relational framework. Rather, I propose it as a refinement: a way of specifying the conditions under which certain relational facts become stable and shareable across systems. RQM, as it stands, leaves underspecified the conditions under which a measurement basis is selected or a particular interaction counts as a measurement in the decohering sense. The agent-centric perspective addresses this gap by identifying the physical constitution of the observer as the source of that structure.\footnote{This approach is distinct from Healey's pragmatist interpretation. Whereas Healey locates quantum state ascription in an agent's physical situation, understood as a context for legitimate inferential practices, the account here goes further by treating the agent's constitution as not merely a pragmatic constraint but a constitutive feature of the system--environment decomposition itself. That is, it is not just that the agent's situation affects what they can justifiably say about a system, but that it shapes what decoheres and how. The agent is not external to the ontological process, but structurally intertwined with it.} In this way, the agent does not override relationality but participates in it: agents are physical systems with particular coupling constraints that generate a context in which certain quantum interactions give rise to classical facts. From this perspective, my account can be seen as a special case of RQM -- one in which the observer is not an abstract placeholder for any interacting system, but a system with a structured relation to the environment that grounds decoherence in a principled and embodied way.

Before spelling out more fully the philosophical implications of this view in the next section, it is worth saying a few words here about the relationship between my account and QBism. Whilst this view shares with QBism an emphasis on the central role of the agent in the quantum description, it differs in both orientation and explanatory ambition. QBism treats the quantum state as a purely subjective degree of belief---a tool for managing personal expectations---and avoids attributing structure to the world independently of the agent's beliefs. By contrast, the agent-centric view developed here is not subjectivist in this sense: it holds that the agent's embodied interaction with the world gives rise to stable patterns that are structurally real within the context of that interaction. These patterns are not detached from the agent, nor are they reducible to belief states; rather, they reflect the physical and operational constraints that shape the emergence of classical facts. However, recent work, including \citep{Bitbol2020} and the edited collection \citep{Berghofer2023}, especially \citep{Fuchs2023}, suggest a growing recognition within QBism that the agent's interaction with the world plays a constitutive role in how phenomena arise.

By the same token, the present account departs from QBism in its framing of the agent's embodiment. In particular, this view assumes that measurement interactions are shaped by generalisable features shared by a class of agents, not by the idiosyncrasies of a single agent's physical constitution. As such, instruments function not merely as prostheses tailored to a particular body (as argued in \citep[\S4]{Pienaar2023}), but as apparatuses whose operational regularities derive from coordination with stable features of the `bodily schema' shared by agents like us (more on this below). To this end, objectivity is not merely a by-product of ignoring differences across agents, as QBism might suggest, but arises from a structurally shared orientation to the world. Having said this, I see this issue of intersubjectivity as an emerging problem of significance in quantum foundations, and I suspect there is much more to be said about this.

\section{Embodied phenomenology of quantum mechanics}
\label{sec:embodphenqm}

In this section I explore in more depth the connection between the account in the preceding sections and phenomenological approaches to understanding the world, particularly the embodied phenomenology of \citet{MerleauPonty1945}. As I mentioned in the introduction, I am emboldened to pursue this connection following French's work on phenomenological approaches to quantum mechanics \citep{French2024}, as well as an earlier edited collection on phenomenology and QBism \citep{Berghofer2023}. In particular, \citet{French2024a} argues for a phenomenological synthesis of QBism and RQM via \citepos{Massimi2022} framework of perspectivalism realism. Whilst French appeals to the reflective act to explain how objectivity is constituted within a shared lifeworld, the present account aims to provide a complementary physical framing by tracing how decoherence stabilises facts relative to the embodied constraints of agents. As such, the present view is less concerned with resolving philosophical tensions in realism, and more focused on explaining the structural role of agency in grounding classicality.\footnote{One more comment about intersubjectivity: Whilst the present account does not appeal to intersubjectivity as a prior condition for objectivity as French does, it does assume that decoherence stabilises observables relative to structural regularities across a class of embodied agents. In this sense, intersubjectivity is not foundational, but it is a natural consequence of the shared constraints that structure agents' bodily orientations and sensorimotor couplings with the world. The effective system-environment split is therefore fixed not by individual perceptual habits, but by the general conditions under which stable classical information becomes accessible to agents like us. This contrasts with QBism, where objectivity arises only through post-hoc alignment of personal experiences. Here, intersubjective convergence is built into the dynamics that shape fact-formation from the outset.} Let us consider this view in a bit more detail.

In the view I have presented here, the preferred basis that emerges in any given measurement interaction is not simply imposed by an objective environment that is in some sense `out there' in the world, but rather reflects the physically and structurally conditioned possibilities of the observing system -- especially when that system is a human agent. As I argued in \S\ref{subsec:agcentobj}, in those cases where the observing system is some measuring device, those systems too are specifically designed to concatenate with our own idiosyncratic physical constitution -- they are \emph{our} instruments. Thus, our measuring apparatuses, modes of interaction, and then the very variables that are accessible to us to objectify are shaped by our physical constitution. Most importantly in this context is that this does not mean that the preferred basis that emerges in some interaction between agent and world is somehow `chosen' subjectively, far from it; the space of possible interactions is bounded by the forms of interactions our bodies and instruments permit, just as they are too constrained by the `worldly infrastructure' of the physical systems.

This account has some striking similarities with Merleau-Ponty's phenomenological program, particularly his insight that perception is always structured by an embodied mode of access to the world. At the core of Merleau-Ponty's program is a rejection of the Cartesian dualist paradigm in which the thinking mind is distinct from the extended body. Merleau-Ponty argues that our perceptual access to the world is fundamentally embodied, and that our lived bodily experience denies any absolute split between subject and object, mind and body. In this way, he maintains that our bodily being is the centre of perspective and agency through which the world is disclosed. As such, according to Merleau-Ponty, agents are incapable of perceiving the world from a disembodied perspective, but instead do so always from an embodied orientation that simultaneously constrains but also opens the kinds of phenomena that can appear -- ``the body is our general medium for having a world'' \citep[p.169]{MerleauPonty1945}.

In this context, Merleau-Ponty talks of the `body schema', which is not a reflective image of one's own body but a system of sensorimotor capacities, possible actions, and perceptual orientations common to embodied experience, and so underpins perception as an operative structure that coordinates inter-sensory couplings and governs an agent's engagement with the world: ``To have a body is to possess a universal setting, a schema of all types of perceptual unfolding and of all those inter-sensory correspondences which lie beyond the segment of the world which we are actually perceiving'' \citep[p.381]{MerleauPonty1945}. As such, the perceptual world is not constituted privately, but emerges through a structural coupling between the body and the world---``My body is the fabric into which all objects are woven'' \citep[p.273]{MerleauPonty1945}. On this basis, Merleau-Ponty insists that it is through the body schema that ``the unity of the body'', and ``the unity of the senses and of the object'' come into view \citep[p.273]{MerleauPonty1945}. To this end, my use of the term `agent' throughout this work should be understood as a member of a class of embodied observers who share broadly similar physical and perceptual structures, consistent with Merleau-Ponty's notion of the body schema. Moreover, this framework supports the interpretation advanced above, that classical facts are stabilised not for a solipsistic observer, but for structurally similar embodied agents.

Importantly for Merleau-Ponty's program, he attempts to identify the pre-reflective layer of experience that exists prior to the abstraction of the world into subject and object. According to his view, perception is not a passive recording of sensory data nor a detached intellectual judgment but an active engagement of the embodied subject with its surroundings. In perceiving, the world is immediately experienced as meaningful through our bodily orientation: everything we perceive is imbued with significance relative to our embodied situation. As part of this, Merleau-Ponty argues that subject and object, mind and world, are not irreconcilable opposites but interdependent facets of one continuous experience.

There is a clear sense, then, in which the agent-centric constraints outlined above on the emergence of a preferred basis in measurement can be seen, in Merleau-Ponty's terms, as a quantum analogue of the role of embodiment in structuring perceptual access. Just as our embodied engagement with the world conditions what can appear as a stable object of experience, so too do the physical structures and limitations of our measuring interactions condition which relational quantum states can be stabilised and rendered epistemically accessible. Thus, just as Merleau-Ponty argues we never perceive objects `in themselves' but always in relation to our bodily capacities, if the above line of argument is along the right track than what becomes actualised or stabilised as a `fact' in the process of quantum measurement depends on how the measuring system---which includes us---is capable of interacting with the physical quantum system. Importantly, the view here is not that agents freely choose a basis; rather, the effective basis is the upshot of how agents like us are embodied and instrumentally coupled. In Merleau-Ponty's dialectic of subject and object \citep[IV]{MerleauPonty1942}, such constraints are not external add-ons but the very conditions under which facts can become stable \emph{for us}.

This phenomenological orientation is highly consistent with Michel Bitbol's program, which reads quantum mechanics through the lens of situated embodiment and rejects representational `views from nowhere'. Bitbol emphasises that quantum theory is first and foremost a rule for organising an agent's situated expectations and interventions, and he relates this stance explicitly to Merleau-Ponty's endo-ontology and to QBism's participatory turn \citep{Bitbol2020,BitbolDeLaTremblaye2023}. The present account is broadly sympathetic: I take the ontology of quantum facts to be endo-ontological in Merleau-Ponty's sense, while the emergence of stability on my account is eco-phenomenological in Bitbol and de la Tremblaye`s sense. My focus here, however, is not on providing an account of the experience or beliefs of embedded agents, but on providing a physical articulation of \emph{why} certain observables become stably available for agents like us -- by tying basis selection to shared embodied interactions and to decoherence-induced pointer structures. In short, where Bitbol stresses the phenomenological primacy of situation, I supply the corresponding operational mechanism: agent-conditioned decompositions that make decoherence explanatory for a class of observers, thereby yielding facts that are \emph{stable for us}.

The result of all this is that the preferred basis that arises through decoherence is not merely a formal mathematical selection. It is, at least in part, a phenomenological constraint that is rooted in the physical and operational structure of the observer. The classical world we extract from the quantum substrate is thus structured by our embodiment, not just passively received. At the very least, this connection deserves increased attention in the quantum foundations community, building upon work such as \citep{French2024} and \citep{Berghofer2023}.

\section{Final thoughts}

I have argued in this paper for the consistency of a situated, agent-centric, perspectival understanding of decoherence and the preferred-basis problem, established on top of the notion of an agent-centric objectification of the structure in the world around us. Using a recent debate concerning the status of measurement in RQM as a platform into these issues, I propose that there is a fundamental role that the agent plays in bringing about the quantum world. On this view, classicality emerges from the interaction between agents and quantum systems as a function of the constraints imposed by the agent as well as the structure of the physical system itself. I contend that the emergence of a preferred basis is not some agent-independent universal feature of interactions between arbitrary systems but is defined relative to the kinds of system--environment splits that are natural for the relevant measuring agent's physical constitution and available measurement resources.

Although I do not claim that my proposed solution to the no-go theorems of Brukner and Pienaar is superior to that offered by Di Biagio and Rovelli, especially not since the proposal I advocate goes beyond the spirit of RQM, if nothing else I hope it provides a viable alternative for understanding measurement as agent-relative in quantum mechanics. Furthermore, whilst in previous work I have made the connection between this kind of view and Bohr's philosophy of quantum theory \citep{Evans20}, I make here the connection with Merleau-Ponty's embodied phenomenology and recent work in quantum foundations in this direction. These two different connections may not be mutually exclusive.

The broader implication of this analysis is that quantum mechanics should not be interpreted as a detached, third-person description of reality. Rather, it should be understood as a structured, agent-relative formalism that encodes the interactive relations between physical systems. The classical world does not emerge universally from quantum mechanics; instead, it emerges on this view relative to equivalence classes of agents with shared physical constraints.

In sum, this paper proposes that the emergence of classicality from quantum systems is not a brute physical fact, nor a purely relational phenomenon, but the result of a structurally constrained interaction between agents and the world. Decoherence, measurement, and the preferred basis are not defined universally, but relative to the physical constitution and operational capacities of the observer. This reframing both clarifies the limits of RQM and offers a philosophically grounded path forward: one in which quantum mechanics is neither the universal language of nature nor the projection of subjective belief, but a structured expression of embodied interaction. In this sense, quantum theory becomes not less real, but more relationally real -- not a detached description of a world from nowhere, but a dynamic account of how the world becomes intelligible to agents like us.

\section*{Acknowledgements}

I am appreciative of useful discussions with Emily Adlam, Gerard Milburn, and Jacques Pienaar. I acknowledge support from the University of Queensland and the Australian Government through the Australian Research Council (DP250102035).

\providecommand{\noopsort}[1]{}

\end{document}